# Advancing glaucoma research with multiphysics continuum mechanics modelling: Opportunities and open challenges.

Daniel Sebastia-Saez[1], Jinyuan Luo[2], Mengqi Qin[2], Tao Chen[1], Cynthia Yu-Wai-Man[2]

1. School of Chemistry and Chemical Engineering, University of Surrey, Guildford, GU2 7XH, UK.
2. Faculty of Life Sciences & Medicine, King's College London, London, SE1 1UL, UK.

**Corresponding Authors:**

Cynthia Yu-Wai-Man, King's College London, cynthia.yu-wai-man@kcl.ac.uk

Daniel Sebastia-Saez, University of Surrey, j.sebastiasaez@surrey.ac.uk

## LIST OF ABBREVIATIONS

| | |
|---|---|
| AGV | Ahmed glaucoma valve |
| AI | Artificial intelligence |
| CFD | Computational Fluid Dynamics |
| ECD | Endothelial cell damage |
| FDA | US Food and drug administration |
| FSI | Fluid-solid interaction |
| GDD | Glaucoma drainage device |
| IOP | Intraocular pressure |
| LPI | Laser peripheral iridotomy |
| MD | Molecular dynamics |
| MIGS | Minimally invasive glaucoma surgery |
| OCT | Optical coherence tomography |

**Abstract**

This review examines the emerging role of mechanistic mathematical models based on continuum mechanics to address current challenges in glaucoma research. At present, the advent of Artificial Intelligence and data-based models have resulted in significant progress in drug candidate screening, target identification and delivery optimization for glaucoma treatment. Physics-based models on the other hand offer mechanistic insight by modelling fundamental physical knowledge. Mechanistic models, and specifically those based on continuum mechanics, have the potential to contribute to a better understanding of glaucoma through the description of intraocular fluid dynamics, mass and heat transfer and other basic physical phenomena. So far, these models have expanded our understanding of ocular fluid dynamics, including descriptions of fluid flow profiles within the anterior chamber of the eye under glaucomatous conditions. With the ongoing development of multiphysics modelling frameworks, there is increasing potential to apply these tools to a wide range of current challenges within the field of glaucoma research. These challenges include glaucoma drainage devices, minimally invasive surgical procedures, therapeutic contact lenses, laser-based interventions like peripheral iridotomy, and the design and optimization of biodegradable drug-releasing intracameral implants, which support patient-specific strategies for glaucoma diagnosis and treatment.

## 1. Introduction

Glaucoma is the leading cause of irreversible blindness worldwide (Tham et al., 2014). It is characterized by imbalanced intraocular pressure (IOP) caused by impaired aqueous humour outflow as illustrated in Figure 1 (Weinreb et al., 2014; Gabelt and Kaufman, 2005). Imbalanced IOP is a major risk factor for glaucoma (Vroemen et al., 2019). In recent years, the integration of computational science to ophthalmic research has greatly contributed to transform how glaucoma is studied. Computational science offers multiscale insights that are ethically and/or technically difficult to achieve through animal experimentation, hence reducing the need for animal testing in the current landscape of ever more restrictive pharma regulation. So far, data-based simulations appear to have had greater impact on glaucoma research than mechanistic ones. For instance, (M. J. Kim et al., 2023) mainly focused on data-based simulations for glaucoma research in their review article, specifically those leveraging machine learning for pattern recognition in clinical datasets. The application of data-based models and specifically the advent of Artificial Intelligence (AI) have caused a surge in the analysis of vast amounts of clinical data, resulting in increased detection rates, enhanced understanding of the causes of the disease, and the design of improved treatment strategies (Huang et al., 2023). However, such models and correlations often lack mechanistic insight. This limitation underscores the complementary role of physics-based (also known as mechanistic) models as opposed to data-based models (also known as statistical models). Mechanistic models are models based on the equations that describe the underlying physical phenomena, such as the equations of fluid flow, tissue biomechanics, and molecular interaction. Physics-based modelling is claiming its place in the digital transformation of glaucoma research in recent years because unlike purely statistical models, they enable hypothesis testing of causal mechanisms (Dvoriashyna et al., 2023). On the downside, extensive validation is required against experimental or clinical data (Nguyen and Ethier, 2015). Mechanistic model is an umbrella term for models that solve the mathematical equations that describe natural phenomena at different length and time scales, each of which with different applications in glaucoma research. At the smallest scale (i.e., nanometres and nanoseconds), Molecular Dynamics (MD) provide insight into the interaction of drugs and the eye structures at the molecular level, for example how a drug molecule interacts with the molecular structure of the trabecular meshwork during a treatment. MD simulations rely however on idealized assumptions about protein structures and solvent conditions, which limit their predictive power.

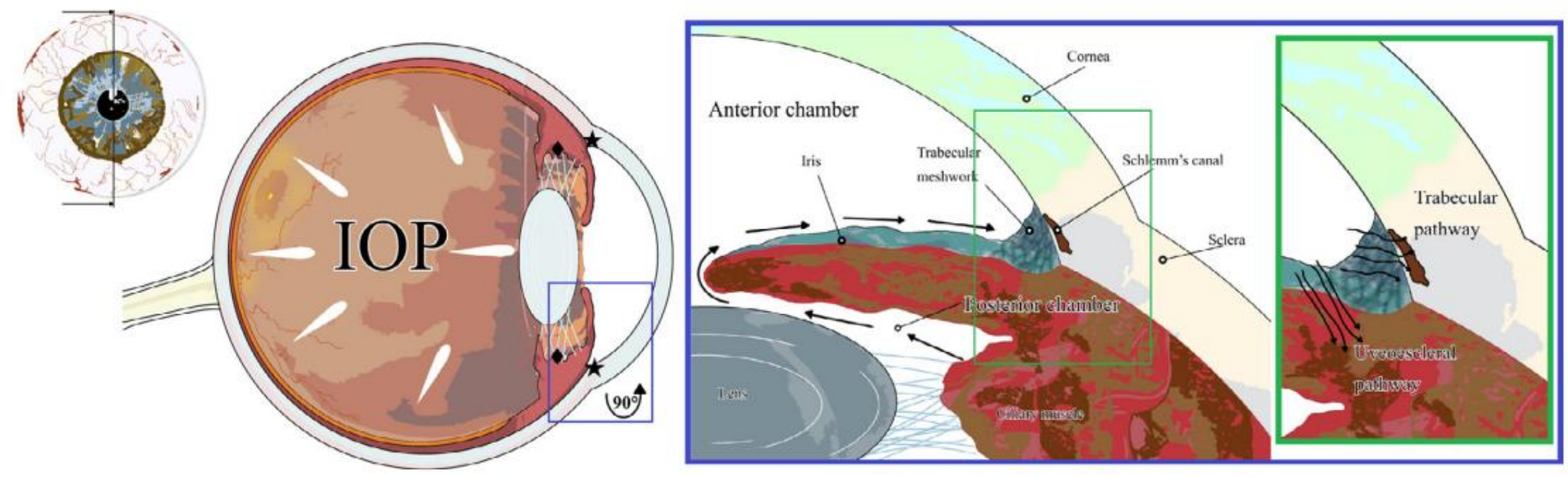

*Figure 1. Schematic illustration of the flow of aqueous humour in the eye's anterior chamber and the pathogenesis of glaucoma. The aqueous humour is produced in the ciliary body and flows around the iris towards the trabecular meshwork. Pressure built up when the trabecular meshwork loses its capability to drain the aqueous humour is transmitted towards the rear part of the eye, hence damaging the optic nerve. Reproduced with permission from (Martínez Sánchez et al. 2020).*

Mechanistic models based on continuum mechanics provide an appropriate framework for modelling ocular phenomena at larger space scales when the characteristic length scale exceeds the mean free path of the particles, as continuum mechanics models solve the differential equations for the conservation of mass, momentum and energy in solids and fluids instead of molecular interactions. Continuum mechanics provide descriptions of bulk behaviour allowing quantitative description of intraocular fluid circulation (i.e., the aqueous humour), non-Newtonian fluid movement (i.e., the vitreous), scleral and corneal deformation, solute transport in drug delivery and thermal effects in the eye, including aqueous humour convective movement or quantification of external radiation, among other specifics. Continuum mechanics models for the description of flow fields make use of a set of techniques called Computational Fluid Dynamics (CFD), the use of which has seen a sharp increase in Biomedical Engineering in the past years. Early applications of continuum mechanics simulations in glaucoma research primarily emphasized fluid domains in isolation as will be seen in Section 2, (e.g., the anterior chamber without considering the vitreous), often neglecting the mechanical feedback from surrounding ocular tissues. However, glaucoma pathophysiology demands a multiphysics modelling approach, including fluid-structure interaction (FSI) between the aqueous and the vitreous humour and the surrounding hyperelastic tissues (Drobny and Friedmann, 2022). Also, despite recent advances in multiphysics approaches, current models include assumptions such as isotropic physical properties, homogeneous tissue properties or idealised geometries, which render their applicability difficult to ascertain in real-life scenarios (Karimi et al., 2025; Avtar and Srivastava, 2007).

Mechanistic understanding of intraocular fluid dynamics, as well as the interplay with the deformation of the eye's biological tissue is crucial to advance the development of effective IOP-reducing therapies. Alongside ethical restrictions, direct experimental measurements of intraocular processes at the millimetre scale remains highly challenging, resulting in excessive difficulties to place invasive sensors to obtain data on the pathogenesis of glaucoma. In this context, mathematical modelling can contribute to significant advances in the field by generating mechanistic hypotheses, interpreting existing experimental data sets, and testing therapeutic interventions for which it would be otherwise difficult to obtain preliminary evaluation.

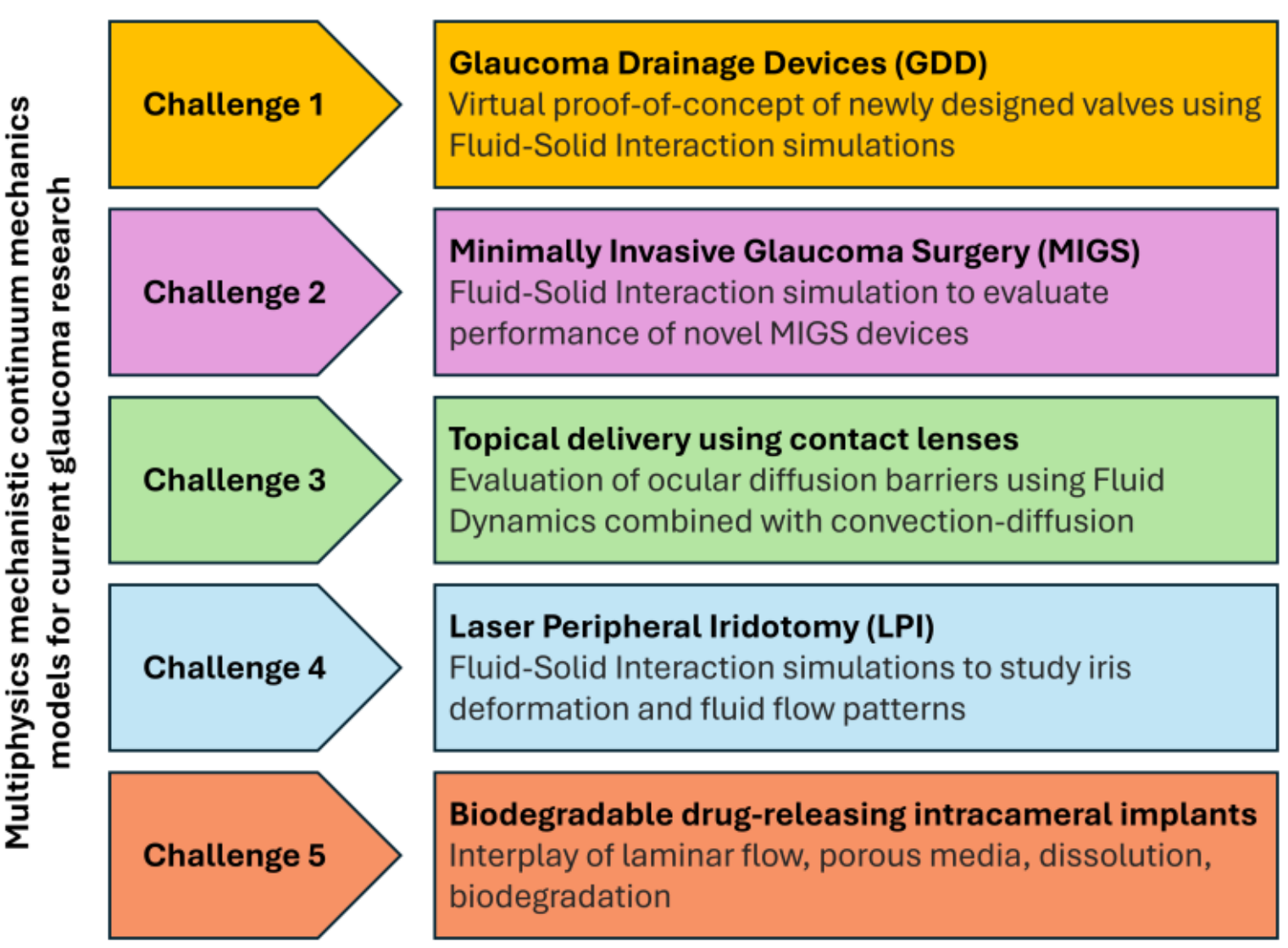

*Figure 2. Schematic showing current challenges in glaucoma research and how multiphysics continuum physics simulations can help.*

Significant challenges remain unmet in the field of glaucoma research, summarised in Figure 2. Trabeculectomy has been the mainstay of surgical treatment for medically uncontrolled glaucoma for several decades (Cairns, 1968), but carries the risk of serious complications, including postoperative hypotony, bleb scarring, and infections (Rao and Cruz, 2022; Razeghinejad et al., 2017). Trabeculectomy shortcomings have led to a rising interest in the development of minimally invasive glaucoma surgeries (MIGS), which carry a lower risk of complications while keeping comparable IOP-lowering effects to conventional surgeries. MIGS are increasingly becoming a front-line glaucoma surgery practice due to several advantages with respect to common treatments (Otarola and Pooley, 2021). In this context, mechanistic modelling at the continuum mechanics scale can play a crucial role in exploring ocular fluid mechanics in different MIGS scenarios by introducing preliminary proof-of-concept, fast prototyping and testing of MIGS devices in an animal-free manner. Other solutions to current glaucoma research challenges include glaucoma drainage devices (GDD) (Garg et al., 2025), topical drug delivery using contact lenses (Franco and De Marco, 2021), laser peripheral iridotomy (LPI) (Rifada et al., 2025), and biodegradable drug-releasing intracameral implants (Tsung et al., 2023). The development of these therapeutic strategies is nowadays constrained by incomplete understanding of ocular fluid dynamics. Continuum mechanics modelling provides a framework to test device-tissue interactions, quantify the effect of biological variability on surgical procedures and optimize designs before clinical testing. Although there has been an increased degree of complexity over recent years, existing models still oversimplify tissue anisotropy and other aspects like patient-specific geometries, hence limiting their applicability. Filling these gaps will be crucial to fully integrate mechanistic models into a reliable virtual environment for device and procedure innovation.

With the development of complex multiphysics capabilities in continuum mechanics mathematical modelling software, the mechanistic understanding of glaucoma pathophysiology is expected to increase in the future towards the development of novel and efficient therapies. One of these multiphysics capabilities is fluid-structure interaction (FSI), some applications of which are the study of the interaction between aqueous humour, the trabecular meshwork and Schlemm's canal stiffness to understand IOP regulation (Wang et al., 2017) and corneal deformation caused by puff tonometry (Maklad et al., 2020). Novel computational mechanistic methodologies in the field of ophthalmology are not restricted to FSI applications. CFD has also been coupled to heat transfer to study convective currents in the flow of aqueous humour, which present changes depending on body position (Tang et al., 2022). Finally, coupling CFD with species transport has resulted in the description of drug delivery mechanisms (Pandey et al., 2024).

The body of literature describing these applications is scarce at present and often constrained by oversimplifying assumptions. The advent of increasingly powerful computers sets the stage for mechanistic ocular digital twins capable of substituting animal experimentation in the future. Ongoing research offers a glimpse into a future where computational methods play an ever-increasing role in glaucoma research. This review shows the state-of-the-art of mechanistic models for glaucoma research at the continuum mechanics scale level and how they can help in the future with the current challenges in glaucoma research.

## 2. General trends in the use of mechanistic continuum physics models in glaucoma research: From CFD to a comprehensive multiphysics integration.

Continuum mechanics models have led to an enhanced understanding about the flow field in the anterior chamber and the vitreous humour. Some models dealing with the description of ocular fluid dynamics are summarised in **Error! Reference source not found.**, both using CFD only and FSI simulations.

*Table 1. A list of relevant literature on continuum mechanics modelling of the aqueous humour and vitreous humour.*

| Reference | Summary |
|---|---|
| (Martínez Sánchez et al., 2020) | OCT-derived 3D geometry CFD model of aqueous humour flow with detailed drainage system (trabecular meshwork, Schlemm's canal, collector channels). Simulated pathological cases (narrow angles, blocked drainage), where velocity fields showed altered pressure and velocity patterns. |
| (Fitt and Gonzalez, 2006) | A pioneer article on the mathematical treatment of the aqueous humour flow in the anterior chamber. Does not include important aspects of flow in the anterior chamber, such as trabecular outflow resistance. |
| (Villamarin et al., 2012) | Realistic geometry based on detailed histological data. Pressure and velocity profiles obtained to assess impact of glaucoma on IOP and effectiveness of trabeculectomy and glaucoma drainage devices (GDD). |
| (Murgoitio-Esandi et al., 2023) | The model was used to investigate the link between IOP and iridocorneal angle closure. Presents a simplified geometry compared to true patient anatomy. The common assumption of Newtonian, steady-state aqueous humour is taken. |
| (Repetto et al., 2010) | Included effects of eye motion to simulate saccadic movements, hence obtaining information on how saccades alter fluid flow within the eye. Included, however, Newtonian rheology and rigid boundaries. |
| (Jen et al., 2020) | Described the flow specifics of Descemet membrane detachment. Included a simplified geometry and no tissue mechanics. |
| (Karimi et al., 2025) | Used high-resolution 3D imaging methods to characterise the trabecular meshwork, hence constituting a step forward towards personalised modelling and glaucoma treatment. |
| (Gozawa et al., 2021) | Intraocular gas–fluid interfaces after vitrectomy were modelled using CFD to estimate how buoyancy and ocular geometry affect shear stress on the retina. Simplifying assumptions include the treatment of vitreous substitutes as Newtonian fluids and idealized eye geometries. |
| (Silva et al., 2020) | Vitreous flow during saccadic eye movements was described using CFD. The model was used to compare intact viscoelastic vitreous with liquefied vitreous description. Relevant for the study of ocular aging. |
| (Ong et al., 2024) | CFD model complemented with 3D confocal image of rat retinal vasculature to study how changes in vascular pressure affect retinal wall shear stress distributions. Animal vasculature used instead of human |
| (Avtar and Srivastava, 2007) | Aqueous humour outflow through the trabecular meshwork was modelled |
| | using simplified porous-media and annular representations to estimate resistance. |
| (Yan et al., 2022) | Fluid-structure simulation of different iris-lens channel distances upon diabetic conditions. |
| (Ariza-Gracia et al., 2018) | Comparison of modelling approaches to simulate corneal non-contact tonometry. Not including the description of corneal mechanics leads to overestimated corneal displacement. |
| (Ferroni et al., 2020) | Evaluation of the interaction between the vitreous humour and a magnesium-based device for the treatment of glaucoma and the comparison between Fluid-Structure interaction and fluid dynamics description. Minor differences found between both methods. |
| (Maklad et al., 2020) | Presents a model of the human eye under air-puff tonometry using a coupled CFD and finite element approach to capture fluid–structure interaction. Simulations reproduce corneal deformation and pressure profiles. |
| (Sassetti et al., 2011) | FSI analysis of movable valve to change hydraulic resistance and to adjust outflow according to variations in IOP. |
| (Wang et al., 2016) | Descriptive study of the deformation of the iris upon changes in IOP. Numerical methodology validated using Particle Image Velocimetry. |

All the literature gathered in **Error! Reference source not found.** show that the application of CFD to the description of fluid flow in the anterior chamber and the vitreous humour is still currently in development since the first published work in the early 2000s. This approach has been key in the understanding of the general fluid flow (Figure 3D), detailed descriptions of the trabecular meshwork (Figure 3A and B), and multiphase flow in the vitreous (Figure 3C). A high-fidelity description of the geometry obtained with the aid of imaging methods opens the door to personalised studies. However, these studies lack the accuracy of FSI simulations, which allow the study of specific applications of the interaction of the eye's fluid with the surrounding hyperelastic tissue.

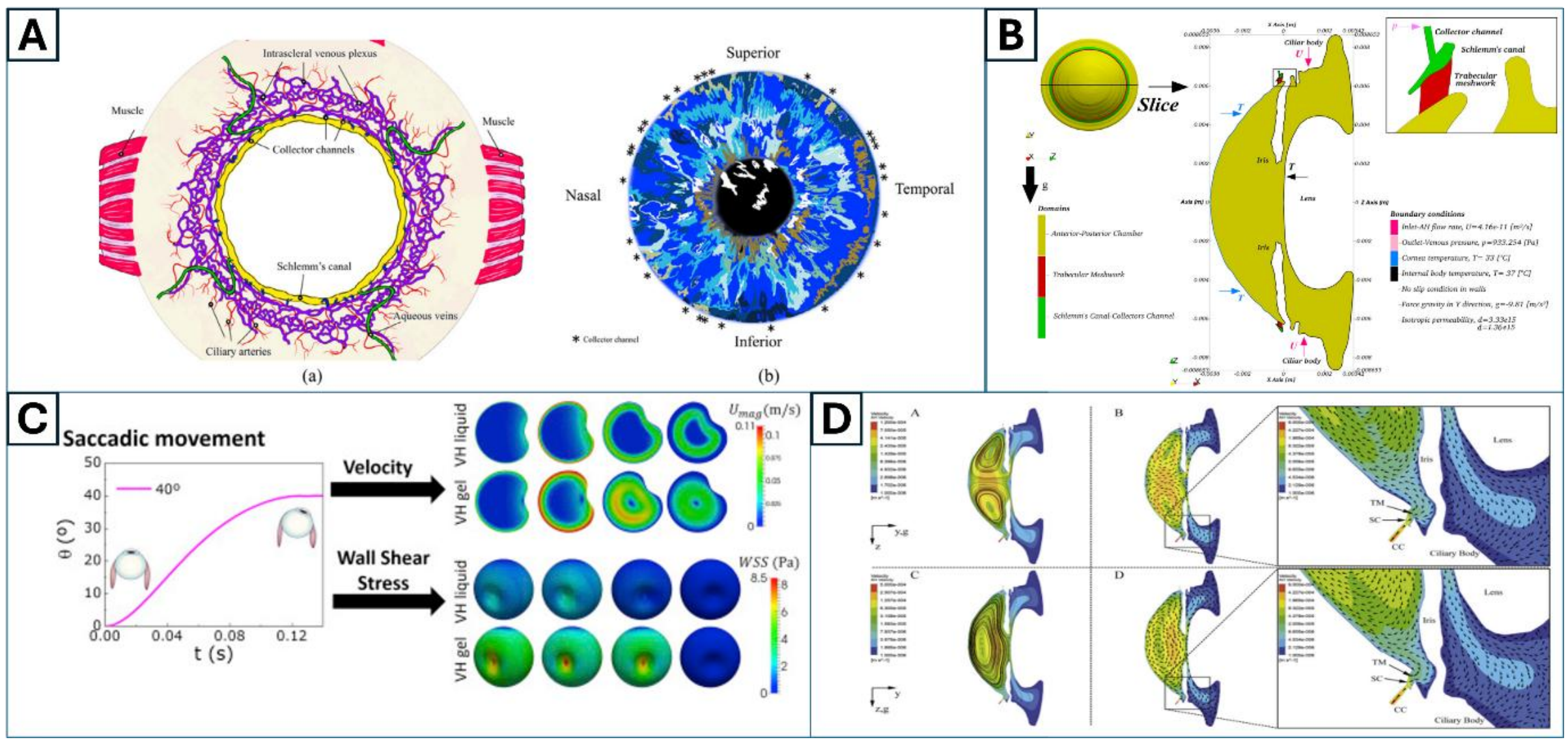

*Figure 3. This figure describes the capabilities of CFD simulations of the eye's fluids, including the aqueous humour and vitreous humour: (A) a schematic of the anatomy of the collector channels in the trabecular meshwork, the simulated geometry is reproduced on the top left image, including a zoomed detail (B) on the top right (Reproduced with permission of Martínez Sánchez* et al., 2020). *Image (C) shows the effect of saccadic movements on the wall shear stress on the eyeball (Reproduced with permission of Silva et al., 2020). Image (D) shows patterns of aqueous humour fluid flow in the anterior chamber in supine and standing body positions (Reproduced with permission of Villamarin et al., 2012).*

The body of reported work on CFD in the field of glaucoma shows an ample range of applications from the description of the fluid field in the vitreous humour under saccadic movements to the evaluation of the fluid dynamics of sensors for IOP monitoring (Seyedpour et al., 2023) and others as described in the beginning of this section. Combining fluid flow with other physics expands the range of applications considerably. For example, fluid-solid interaction (FSI) simulations have been reported describing the effect of fluid flow on the biomechanical properties of the cornea (Maklad et al., 2020). Studies like (Bennett et al., 2023), who use CFD to study the optimisation of postoperative positioning for corneal transplantation patients to enhance surgical outcomes could greatly benefit from FSI approaches. The review by (Nguyen and Ethier, 2015) is particularly interesting and provides insight on how the application of mathematical modelling of the solid mechanics of the eye, and its coupling with imaging techniques can accelerate glaucoma treatment through the development of improved diagnostic tools, IOP control strategies and novel therapeutic strategies. CFD is also useful to study the flow around phakic lenses (Fernández-Vigo et al., 2018; Kawamorita et al., 2012, 2016), to study the effect of heat exposure in the eye (Sharon et al., 2008), and to predict drug release profiles from intraocular lenses, when combined with the equations that describe diffusion mass flow (Clasky et al., 2022). The behaviour of tamponades, such as silicone oil (Rossi et al., 2022, 2021; Wang et al., 2022), and the specific positions that optimise gas tamponades to improve surgical outcomes (Gozawa et al., 2021;

Noohi et al., 2015) have also been studied using CFD simulations. Moreover, CFD has greatly enhanced our understanding of the impact of saccadic movements and fluid dynamics within the vitreous (Ferroni et al., 2018; Silva et al., 2020). Recent developments in modelling have also significantly advanced knowledge of the drug distribution in the posterior eye segment (Jooybar et al., 2014; Lamminsalo et al., 2020, 2018; Missel et al., 2010, 2019) by including mechanistic descriptions of mass transport. CFD simulations can also be combined with experimental techniques, such as adaptive optics scanning laser ophthalmoscopy, revealing significant differences in key parameters between diabetic and non-diabetic eyes, and highlighting the potential of this methodology for early detection and monitoring of diabetic retinopathy (Bernabeu et al., 2015; Lu et al., 2016). Furthermore, ocular drug delivery is another field of study where CFD has greatly contributed to develop knowledge on the diverse barriers in the eye. Computational modelling of drug delivery across the outer blood-retinal barrier can be used as a predictive tool for drug transport across the outer blood-retinal barrier (Davies et al., 2020).

Studies found in the literature based on the description of the flow field in glaucoma applications can be localised in a specific area of the anterior chamber or can combine areas involving different physical phenomena. For instance, a study focused on the fluid dynamics of the Schlemm's canal using constant flow rate data as input to their computational geometry (Guo et al., 2021). Another study also combined the laminar flow description of the aqueous humour in the anterior chamber with the description of the trabecular meshwork as a porous medium (Ferreira et al., 2014). In the latter, the permeability of the trabecular meshwork was intentionally changed to mimic pathological scenarios. The effect of a change in the permeability of the trabecular meshwork on the IOP distribution in the anterior chamber could be assessed following this strategy. An expansion of the capabilities of their model included how a change in the permeability of the trabecular meshwork depending on the applied therapeutic drug could be useful to assess the effectiveness of drug-releasing implants in the future. A similar study using CFD with an idealised geometric model compared normal and glaucomatous conditions using Darcy's law to model the trabecular meshwork (Basson et al., 2022). Fluid flow both in porous media and free flow was also implemented to study how changes in the geometry of a bleb can affect the treatment of glaucoma (Gardiner et al., 2010). These studies demonstrated the potential of CFD to inform and optimize surgical treatments, predict post-operative complications, and tailor procedures to individual patient ocular geometries. Recent studies have utilized CFD to model the biomechanics of the aqueous outflow pathway, constructing detailed 3D models of the eye's microstructure, including the trabecular meshwork and Schlemm's canal (Karimi et al., 2023, 2022). These simulations revealed that increased tissue stiffness in glaucoma leads to higher outflow resistance and altered flow patterns, contributing to elevated IOP. Furthermore, comparative analyses between normal and glaucomatous eyes have shown significant differences in the biomechanical behaviour of the outflow tissues, highlighting the critical role of tissue stiffness in regulating IOP. Another example of multiphysics approach encompasses the use of bulk fluid motion combined with heat transfer and species transport. An example of such a multiphysics approach in the field of ophthalmology has been reported, in which the authors extracted conclusions on how aging affects temperature and drug distributions in the anterior chamber, hence leading to impeded treatments for elderly patients (Bhandari et al., 2020).

In conclusion, the usefulness of CFD multiphysics continuum mechanics simulations in ophthalmology has been proven over the years, with a range of applications that can significantly increase when combined with mechanistic knowledge on the multiphysics phenomena taking place in the eye and/or experimental techniques, offering insight into the pathophysiology of glaucoma and the potential for developing targeted therapies. The body of literature often focuses on the description of the flow field within the eye, with substantial efforts towards the recreation of an accurate geometry of the eye with the aid of imaging methods. Recent trends also show that coupling CFD descriptions of the flow field within the eye with the solid mechanics of structures, such as the iris, lens, trabecular meshwork, and other physical phenomena, open interesting research avenues.

## 3. Current challenges in glaucoma research: What can mechanistic continuum physics models do?

Glaucoma research requires an interdisciplinary approach that includes knowledge on biomechanics, mass transport, flow in porous media, and other phenomena to address specific therapeutic challenges. These key challenges include optimizing glaucoma drainage devices (GDDs) to improve suboptimal aqueous humour outflow while reducing endothelial cell damage (ECD) (Basson et al., 2024), gathering simulated data for minimally invasive glaucoma surgery (MIGS) (Kudsieh et al., 2020), accurately modelling ocular diffusion barriers for topical drug delivery via contact lenses (Wang et al., 2023), optimizing laser peripheral iridotomy (LPI) procedures to minimize excessive and unnecessary stress in surrounding tissue (Cai et al., 2022), and prototyping and testing biodegradable drug-releasing intracameral implants to achieve controlled degradation and delivery profiles (Ioannou et al., 2023; Paleel et al., 2025; Qin et al., 2024). Multiphysics models integrating fluid dynamics with soft tissue mechanics are essential for advancing device designs, improving drug delivery mechanisms, and enhancing surgical outcomes, thereby driving future advancements in glaucoma treatment. By bridging physical behaviour across scales and domains, these models hold the potential to significantly refine device design, personalize drug delivery strategies, and reduce adverse effects.

### 3.1. Glaucoma drainage devices (GDD)

GDDs are surgically implanted shunts designed to facilitate the outflow of aqueous humour, hence lowering IOP by diverting aqueous humour from the anterior chamber to an external reservoir (Lim, 2022). The design and optimization of GDDs require a deep understanding of the fluid dynamics within the eye, but also the interaction of implanted materials with ocular tissue at different levels including biocompatibility and mechanical stress distribution around the eye. CFD has played an important role in the prototyping, optimization, and preclinical analysis of GDDs, thereby offering a non-invasive method to obtain the flow behaviour, pressure and thermal profiles. It enables accurate predictions of device performance, understanding of fluid-structure interactions, and assessment of potential complications, such as endothelial cell damage (ECD) (Basson et al., 2024). By offering detailed insight into the fluid dynamics within the eye, CFD enhances the development of more effective and safer GDDs for glaucoma treatment.

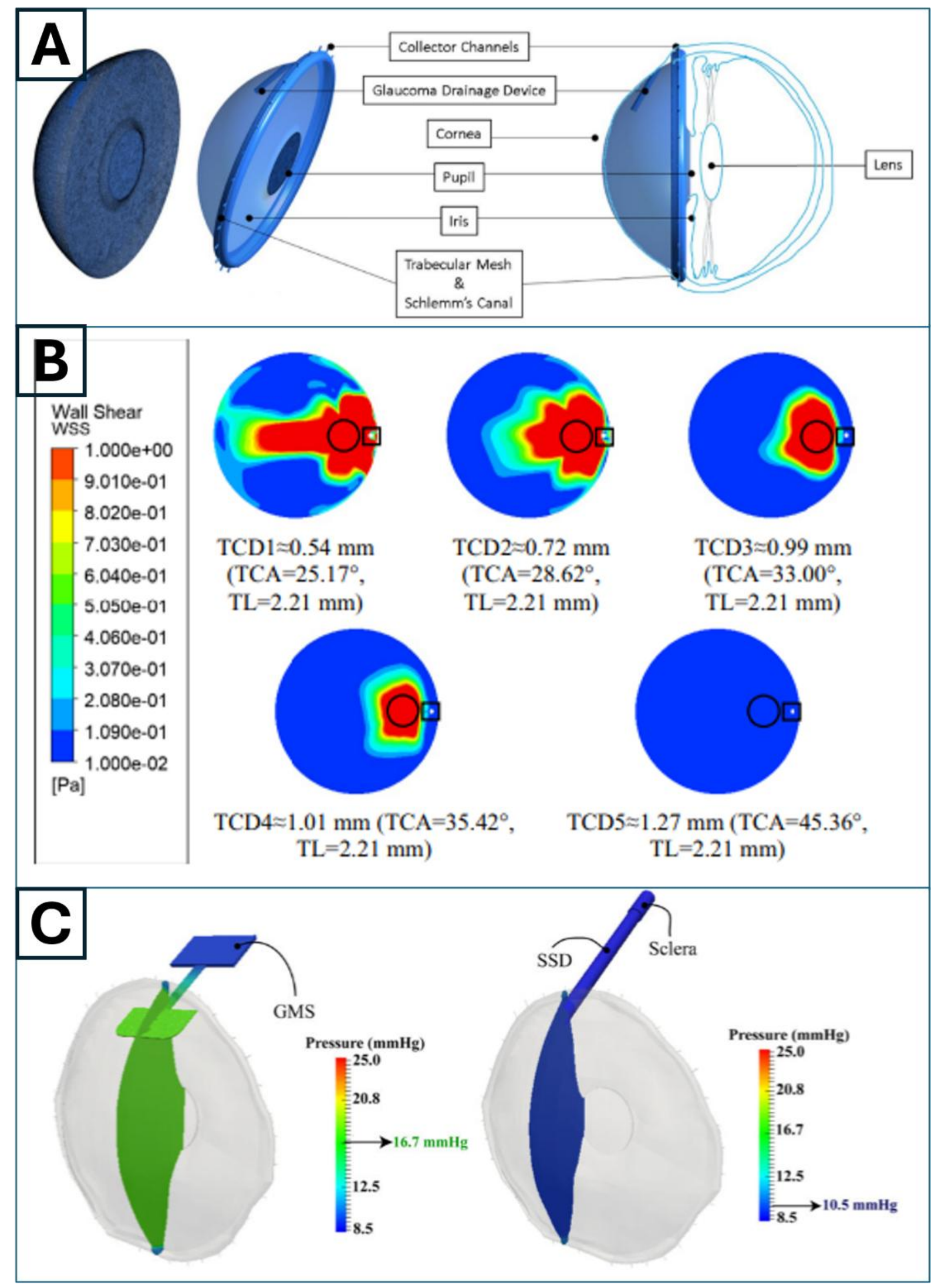

*Figure 4. An illustration of the capabilities of CFD simulations for GDD applications. Image (A): the computational geometry used to study the effects of GDD in ECD. Reproduced with permission from* (Basson et al., 2024)*. Image (B): Colour-coded distribution of wall-shear stress on the cornea, used as a proxy to evaluate ECD. Reproduced with permission from* (Basson et al., 2024)*. Image (C): Results showing the effect of two different GDDs, called GMS and SSD on pressure variation in the eye's anterior chamber. Reproduced with permission from (Mauro et al.* 2018)*.*

CFD models have been used to test GDD prototypes, often as virtual proof-of-concept and limited to the analysis of the flow field to confirm whether or not a specific design can be used to decrease IOP ("Computational and Experimental Analysis of an In Vitro Microfluidic Experimental Setup on Testing Molteno," n.d.)("Computational and Experimental Analysis of an In Vitro Microfluidic Experimental Setup on Testing Molteno," n.d.). Figure 4 provides some illustrative examples on the use of CFD simulations for GDD testing. For instance, a study employed CFD to test the Ahmed glaucoma valve (AGV), identifying flow inefficiencies and proposing new designs with improved hydraulic characteristics (Kara and Kutlar, 2010; Kara et al., 2022). Another study used a multiphysics CFD approach to evaluate how a novel device regulates aqueous humour outflow and stabilizes intraocular pressure (Rafiei et al., 2024). Another study combined fluid flow simulations with experimental observations to characterise the AGV performance (Pan et al., 2005). In their work, the authors compared the fluid dynamics of the AGV and a novel design, revealing that the new model exhibited a non-linear pressure drop response similar to the AGV, while achieving an optimal flow rate and IOP drop. (Mauro et al., 2018) introduced a patient-oriented numerical procedure using CFD to analyze the performance of four GDDs: ExPRESS shunt, iStent inject, SOLX gold micro shunt, and a new silicone shunt device. Another application of CFD is illustrated by (Basson et al., 2024) who investigated endothelial cell damage (ECD) caused by GDDs. The authors developed a CFD model to test the hypothesis that increased shear stress on the corneal wall due to GDD placement could lead to ECD. The findings indicated that maintaining a limited tube-cornea distance and a tube-cornea angle could prevent pathological wall shear stress that may result in ECD. The authors used wall shear stress as a proxy measurement of tissue damage.

Introducing the solid mechanics of the tissue to the description of the fluid flow could provide additional insight on tissue strain, as it is possible nowadays to introduce material failure criteria and tissue damage dynamics in FSI simulations. Such approach would provide a more direct measurement of tissue damage than using wall shear stress as a proxy. Wall shear stress can also undergo changes when considering the interaction between fluid flow and deformable materials, which would require validation of the assumption by comparison between FSI and fluid-only simulation. An evolution towards combining fluid flow with solid mechanics has been observed, where a combination of fluid dynamics and hyperelastic solid mechanics was used to describe how the valve's opening depended on IOP values (Panduro et al., 2021). Transient models combining the fluid flow description with solid motion and deformation within the fluid flow could be used in the future to obtain key performance parameters in terms of fluid flow, and also to check the actual functioning of new GDD designs in a virtual environment. Despite these advances, most published models still rely on rigid-wall assumptions, hence limiting their physiological relevance. Future work should prioritize the integration of validated fluid-solid interaction (FSI), patient-specific material properties and experimental validation to support translation to clinical application of newly designed GDD concepts.

### 3.2. Minimally invasive glaucoma surgery (MIGS)

Minimally invasive glaucoma surgery (MIGS) is a family of novel glaucoma treatment strategies aimed at performing small incisions in the eye tissue to place micro-scale drainage devices, which can help in reducing IOP by increasing aqueous humour outflow (Lee et al., 2020). CFD has proven to be a powerful adjunct in the development and assessment of MIGS devices due to experimental limitations, although research trends observed in the application of CFD in MIGS are similar to those observed with GDDs, where numerical data focused on flow field and not on its interaction with the mechanical properties of surrounding soft tissues. The non-linear, anisotropic mechanical properties of these tissues are important as they can significantly influence surgical outcomes.

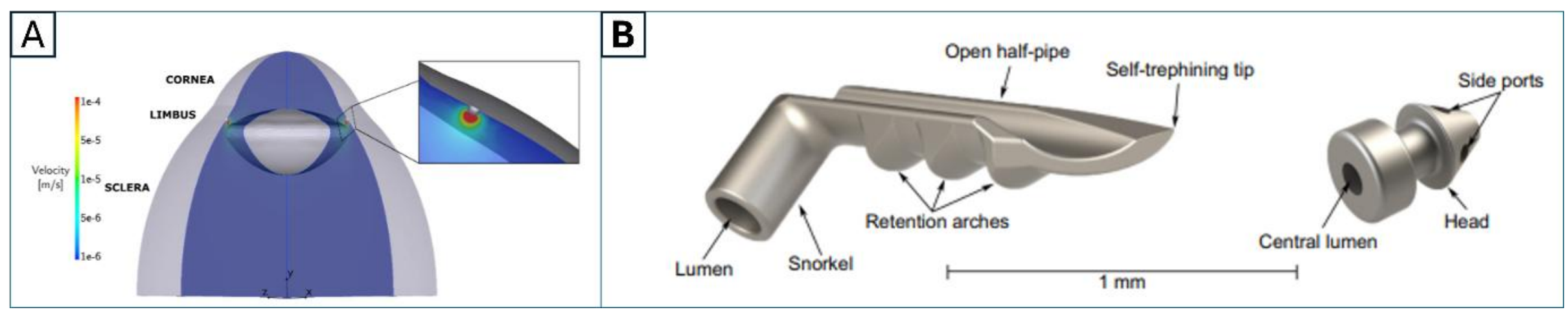

*Figure 5. Image (A): A CFD simulation of the fluid flow in the vicinity of a MIGS device (Reproduced with permission from (Redaelli et al., 2025). Image (B): Illustration of the geometry of two MIGS devices. Reproduced with permission from (Hunter et al., 2014).*

Interesting work has been published on the use of continuum mechanics simulations on MIGS devices, a sample of which is shown in Figure 5. For instance, numerical models were used to compare and predict the efficacy of three MIGS implants: the XEN 45 stent, the XEN 63 stent, and the PreserFlo microshunt (Kudsieh et al., 2020). Implant diameter and filtration bleb pressure were the most critical factors influencing hypotensive efficacy, while tube length and position within the anterior chamber had minimal impact. In another study, *ex vivo* perfusion and CFD simulations were used to analyze the performance of micro-invasive trabecular bypass stents, specifically the Glaukos iStent® and iStent inject®. The results showed that a single iStent significantly reduced IOP, with an additional stent further enhancing IOP readjustment. CFD modelling confirmed smooth laminar flow through the stents at physiological rates, indicating minimal hydraulic resistance and substantial IOP reduction (Hunter et al., 2014). These studies underscore the role of CFD as a powerful prototyping, evaluation and optimisation tool, particularly during the early-stage development of MIGS devices, by enabling controlled and reproducible simulations and parametric sweep over numerous variables, including stent geometry and implantation depth without the ethical and logistic constraints of animal and/or human testing (Kudsieh et al., 2020).

The application of CFD in MIGS device evaluation offers several advantages, such as enabling researchers to isolate and study the effect of various parameters on device performance. Furthermore, CFD provides a detailed, quantitative assessment of fluid dynamics, which is important to understand the mechanisms by which MIGS devices reduce IOP. The non-invasive nature of CFD also allows repeated testing and adjustments to the design of the device without ethical concerns associated with *in vivo* studies. Still, after re-design, newly developed devices must be tested experimentally and ultimately in a clinical environment to ensure its efficacy.

As with the other glaucoma research challenges mentioned in this review, considering fluid-solid interaction can bring advantages for the realistic design of MIGS devices using continuum mechanics simulations. A study (Redaelli et al., 2025) established that models that do not consider fluid-solid interaction underestimate flow velocities compared to FSI models. By including a numerical description of the interplay between the flow field and the hyperelastic mechanical properties of the soft tissue in the area of the trabecular meshwork, hence augmenting the degree of realism, data on the design stage as well as preliminary proof-of-concept could be carried out virtually to advance clinical translation. Despite its benefits, the use of CFD in MIGS device evaluation is not without challenges. The precision of CFD simulations greatly relies on the quality of the input data and assumptions made in model, as the scarcity of high-resolution, patient-specific data restricts the physiological relevance of simulations. Additionally, translating CFD results to *in vivo* conditions can be complex due to the variability in human biology. In this direction, the use of imaging techniques to reconstruct the eye geometry of specific individuals is another step forward towards personalised ophthalmology (Yu-Wai-Man and Khaw, 2016).

### 3.3. Topical drug delivery using contact lenses

Topical delivery by means of contact lenses for glaucoma treatment is a recent drug delivery trend (Yang et al., 2023). From the continuum mechanics standpoint, mass transport of drug solutes used for IOP re-adjustment towards the trabecular meshwork can be described by a combination of the Navier-Stokes equations and the species transport equation complemented with Fick's diffusion and bulk mass transport. Engineering challenges in the development of mathematical models for the design of effective drug releasing contact lenses are mainly concerned with the correct description of the diffusive barriers of the eye, including the lens itself, the tear film, the corneal epithelium, stroma and endothelium, as well as the aqueous humour. A comprehensive description in terms of diffusive and partition properties becomes a limiting step towards the development of fully predictive simulations.

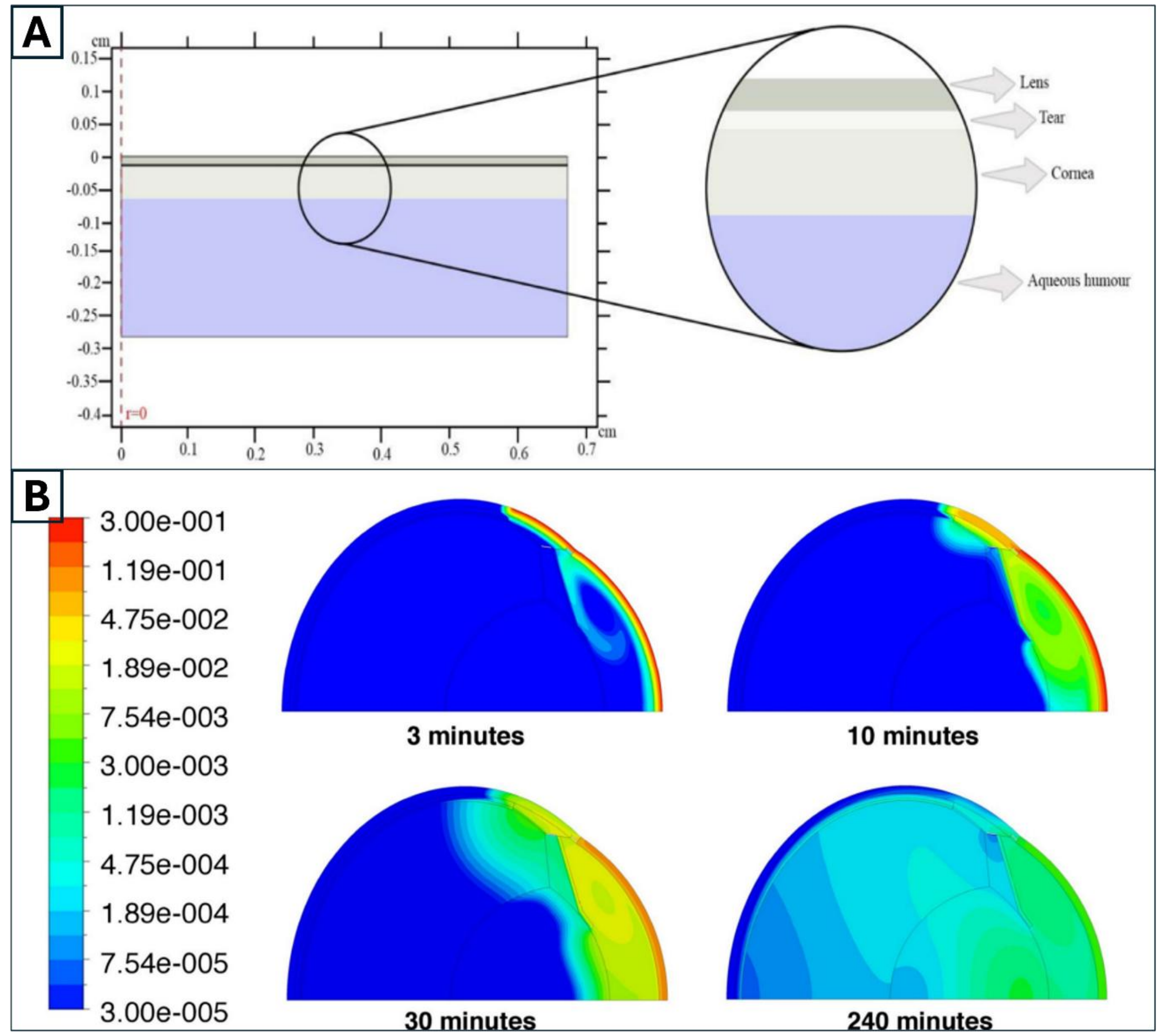

*Figure 6. (A): An illustration of the simplified geometry used by to describe the diffusive barriers encountered in topical drug delivery using contact lenses. Reproduced with permission by (Habibi et al., 2022). (B): Colour map of drug concentration obtained through a CFD simulation of ocular topical drug delivery using eye drops, which presents similarities with the case of therapeutic contact lenses. Reproduced with permission from (Sadeghi et al., 2024).*

The body of literature in the field presents interesting results that highlight the potential of CFD as a tool to study and develop drug delivery through the use of contact lenses. Mathematical and computational models of ocular drug delivery have been recently reviewed, highlighting approaches from compartmental and CFD-based simulations (Sadeghi et al., 2024). The authors pointed out that diffusive barriers are generally treated as simplified resistance terms, with few models explicitly resolving the multilayer cornea or conjunctiva. Their article highlighted the capabilities of current abstractions and the need to capture heterogeneous tissue barriers more faithfully. In a numerical study on topical drug delivery in the eye using contact lenses, a description of the lens, inner tears and cornea in a simplified geometry consisting of static horizontal layers was introduced (Habibi et al., 2022), as shown in Figure 6. Diffusion equations were used to describe the transport of drug from the lens to the aqueous humour, hence not considering convection as the main transport mechanism once the drug reaches the aqueous humour. The simulations were validated against experimental data, which ensures the correct description of the diffusion barrier properties of the lens, inner tears and cornea. The above suggests that at present, the literature regarding the simulation of drug delivery through contact lenses is at the validation stage of the barrier properties in terms of which barriers must be considered along with the adequate values of diffusion and partition coefficients. Other simulation studies for topical drug delivery in the eye have been reported by (Daneh-Dezfuli et al., 2023) and (Yi et al., 2022).

As modelling approaches mature, there is a need to move beyond purely diffusive representations and towards comprehensive multiphysics frameworks. With the correct modelling of the diffusive barriers of the eye, models can be used to tackle engineering challenges concerning the design of therapeutic contact lenses. For instance, regarding the evaluation of the material properties of the contact lens, models can assess whether the use of nanocomposites and polymeric films used in smart contact lenses might affect drug delivery through changes in hydrophilicity, water contents and mechanical properties (T. Y. Kim et al., 2023). Moreover, the evaluation of the mechanical properties of the contact lens through numerical modelling can help with preventing damage in ocular tissues since therapeutic contact lenses must be soft and stretchable to be biocompatible. In this context, continuum mechanics modelling offers a powerful tool for optimizing the pharmacokinetics of therapeutic contact lenses as well as critical device-tissue interactions.

### 3.4. Laser peripheral iridotomy (LPI)

Laser peripheral iridotomy (LPI) is a type of laser therapy to create an opening in the peripheral iris, and to facilitate the outflow of aqueous humour from the anterior chamber to control IOP (Sharif et al., 2024). This artificial channel facilitates the flow of aqueous humour from the posterior to the anterior chamber, bypassing a potentially obstructed pupil. CFD has the potential to contribute towards understanding and optimising LPI and related ocular surgeries, as multiphysics and fluid-structure interaction approaches can shed light on how the geometry of the incisions are affected by the stresses induced by the flow field.

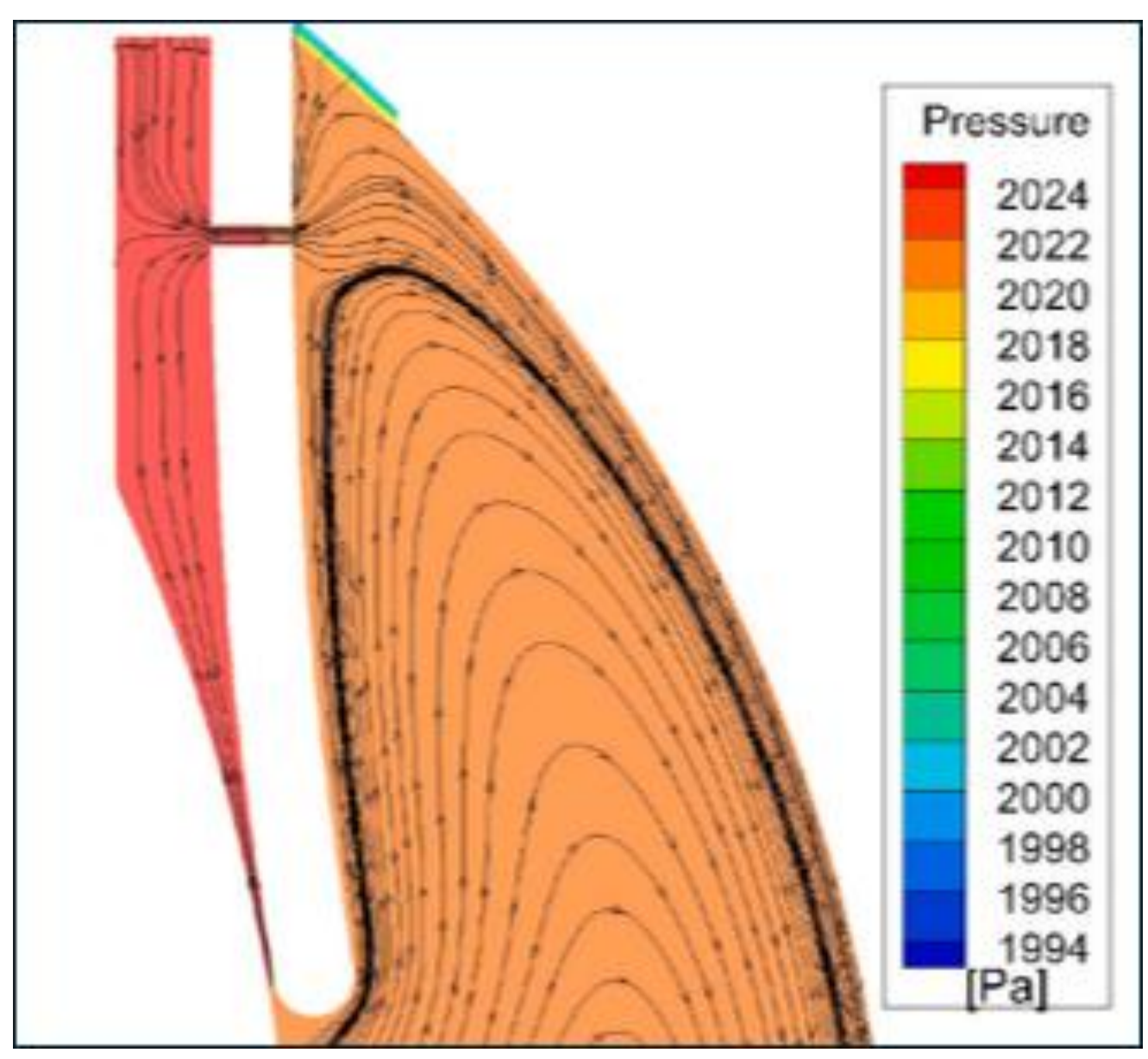

*Figure 7. Illustration of a CFD simulation of an iridotomy. The image shows streamlines and colour-coded pressure distribution for an idealised iridotomy geometry (Reproduced with permission from (Cai et al., 2022).*

In this direction, a recent study utilized CFD to simulate aqueous humour flow and iris deformation under various conditions, including normal and synechial iris configurations. The study found that the pressure difference between the anterior and posterior chambers, along with iris deformation, increased dramatically when the iris-lens gap was smaller than a certain distance. Simulations indicated that an iridotomy could significantly reduce the pressure differential across the iris (Cai et al., 2022). The study however, considered a series of static degrees of iris deformation, although applying solid mechanics balances in the affected tissue has the potential to test the dynamics associated to surgical tissue modification due to invasive procedures, including incision geometry, iris elasticity and the effect of pressure differentials (Figure 7). Their study showed how continuum mechanics models can contribute at present with conceptual insight rather than clinical relevance due to oversimplified anatomy and biomechanics. Also, optical coherence tomography has been used to obtain a high-fidelity ocular geometry, which renders the results more clinically relevant than other idealized geometric approaches (Fu et al., 2025). In the latter, insight was generated on reduced iris-lens contact pressure, which in turn resulted in improved aqueous humour circulation. Their approach allows the implementation of disease-specific anatomy, while still incorporating simplified biomechanics. The majority of the studies reported in the literature had limitations such as idealized geometry, static assumptions and overlooked tissue deformation, which could alter the applicability of the results. Some of these limitations have also been reported by other authors (Chen et al., 2019; Wang et al., 2016; Heys et al., 2001). The range of topics covered also includes the necessity of LPI in the implantable collamer lenses (ICLs) with a central hole, known as hole-ICLs (Kawamorita et al., 2017). The researchers used CFD to simulate the fluid dynamics of aqueous humour in hole-ICLs with and without LPI. The results indicated that hole-ICLs improved the circulation of aqueous humour to the anterior surface of the crystalline lens, reducing the need for further iridotomy procedures. It was concluded that from the perspective of aqueous humour circulation, LPI might be less necessary in hole-ICLs, although they acknowledged the need for further clinical studies to validate these theoretical findings. Their study also highlighted the value of multiphysics applications in optimising surgical procedures for LPI, where an invasive change in the inner geometry of the eye can lead to unexpected stress on the soft tissues induced by artificially modified flow fields.

These studies exemplify how mechanistic continuum mechanics simulations can move beyond descriptive modelling to support surgical decision-making and risk assessment. A current limitation in the current literature, however, is the lack of patient-specific geometry and tissue properties, which constrains the clinical applicability of simulation findings. Future work should prioritize model validation and biological variability towards personalized medicine as well as considering fluid-solid interaction.

### 3.5. Biodegradable drug-releasing intracameral implants.

Biodegradable intracameral implants represent a promising strategy for sustained glaucoma therapy, designed to overcome the limitations of conventional topical and oral drug delivery including most notably poor patient adherence. These implants consist of polymeric scaffolds loaded with active pharmaceutical ingredients, typically prostaglandin analogues such as bimatoprost or travoprost, which are gradually released into the anterior chamber as the material degrades through hydrolysis (Qin and Yu-Wai-Man, 2023). They offer a practical solution to the widespread issue of non-compliance (Moore et al., 2023). Sustained-release systems, once implanted, reduce the burden of daily dosing and minimize the risk of treatment failure due to missed applications.

Their development is however in the early stages, and some concerns exist regarding their functioning. The only FDA-approved example to date is Durysta®, a small cylindrical implant positioned in the iridocorneal angle by means of gravity. Durysta® provides therapeutic drug release for up to 15 weeks. One concern regarding Durysta® is its capability to provide a homogeneous dose in the target tissue. CFD simulations coupled with drug release kinetics and flow in porous media could be a powerful tool to investigate the drug-releasing efficacy of Durysta®. Alternative designs are in development, such as iDose® TR, which uses a trabecular anchor mechanism for fixation, and a novel intraocular lens-based approach from SpyGlass Pharma, in which a ring-shaped drug reservoir is integrated into the haptics of cataract lenses (www.spyglasspharma.com). A similar case of a ring-shaped implant located on an intraocular lens (IOL) has been simulated, including realistic ocular anatomical features and the kinetics of drug releasing from a porous implant (Pandey et al., 2024) shown in Figure 8. Other related studies on porous intraocular implants using a similar multiphysics approach include the work of (Abootorabi et al., 2021), who described drug transport from episcleral implants and studied the effect of implant porosity and position within the eye. Also (Khoobyar et al., 2021) modelled implants as porous capsules but located in the vitreous for the treatment of diseases other than glaucoma, and (Zhang et al., 2017) studied drug delivery from implants located in the anterior segment.

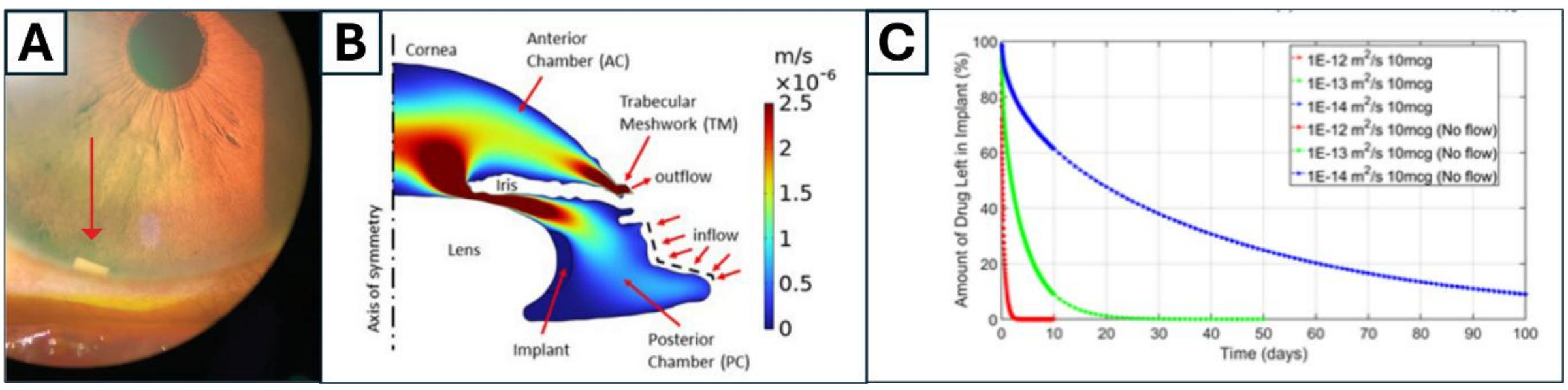

*Figure 8. This figure shows a photograph of the position of the Durysta® implant on the lower part of the iridocorneal angle (A). Image (B) shows a snapshot of the realistic geometry used by (Pandey et al., 2024) as well as the position of the implant in the eye's posterior chamber. Reproduced with permission from (Pandey et al., 2024). The right-hand side image (C) shows the time evolution of the amount of drug left in the implant for 100 days, which was made possible thanks to the multiphysics approach developed by the authors, which included drug-releasing kinetics. Reproduced with permission from (Pandey et al., 2024).*

From a modelling perspective, biodegradable intracameral implants pose a complex multiphysics problem involving fluid flow, mass transport, dissolution and time-dependent material degradation. Continuum mechanics and CFD provide a framework for investigating these coupled phenomena in silico, as models can simulate aqueous humour flow in the anterior chamber and within the porous implant matrix, capturing how shear forces, convective transport, and local concentration gradients influence the implant's degradation profile and drug release kinetics. Moreover, tangential fluid stresses acting on the implant surface may accelerate or spatially vary the degradation process, leading to non-uniform porosity changes and altered drug diffusion paths. Accurately modelling these effects is essential for predicting therapeutic efficacy over time and ensuring consistent drug dosing. In this context, CFD simulations that couple flow fields with dynamic material property changes, such as evolving porosity, diffusivity, and scaffold geometry, can guide the design of the next-generation implants tailored to individual ocular anatomies.

Despite their potential, current models often rely on simplified geometries, constant property assumptions, and lack experimental validation, particularly regarding time-dependent degradation behaviour in vivo. Dynamic changes in releasing rates represent another shortcoming of present models. To advance toward clinically useful predictions, future work must incorporate patient-specific anatomical data, material characterization under physiological conditions, and iterative validation against long-term pharmacokinetic data, as well as further mechanistic knowledge on the kinetics of implant biodegradation.

## 4. Conclusions and future perspectives

Computational methods at the continuum mechanics spatial scale have become an important contributor, although it is still an underutilized tool in glaucoma research. Their strength lies on the potential to provide mechanistic insight that would remain inaccessible to experimental methods due to the ethical and technical barriers. The reported literature has shown that the combination of diverse physical phenomena in simulations can provide abundant benefits in the field of glaucoma research due to the increasing restrictions in the use of animal experimentation and the intrinsic difficulty of experimentation logistics. The majority of reported studies, however, remain restricted to descriptive analysis of the fluid field, often relying on simplified geometries, homogeneous material properties and neglected fluid-solid interaction. This results in constrained predictive power and limit the translation of their results into decision-making knowledge at the clinical level.

In the future, it is envisaged that continuum mechanics simulations will have a crucial role in the current challenges in the field of glaucoma research, including:

1. Glaucoma drainage devices (GDD),
2. Minimally invasive glaucoma surgery (MIGS),
3. Contact lenses for topical drug delivery,
4. Laser peripheral iridotomy (LPI),
5. Biodegradable drug-releasing implants.

To achieve full potential towards contributing to clinical scenarios and personalized medicine, future efforts must shift from oversimplified descriptions towards the following three advances:

1. Integration of multiphysics frameworks capable of capturing coupled processes such as drug transport through dynamic tissues, valve mechanics in GDDs, and implant degradation by combining fluid-structure interaction (FSI) models with diffusion, heat transfer and poroelasticity, towards a fully integrated virtual mechanistic model of the eye.

2. Patient-specific geometries and material data derived from high-resolution imaging and experimental characterization, enabling personalization and improved physiological relevance.
3. Systematic validation strategies, including benchmarking against clinical datasets, to ensure that the models are not only internally consistent but also predictive.

Ultimately, the field must move beyond viewing continuum mechanics models as academic demonstrations and instead embed them into therapy design and device prototyping. Achieving this vision will demand closer collaboration between engineers, ophthalmologists, and industry partners, with the explicit goal of building mechanistic digital twins of the eye that can accelerate innovation while reducing reliance on animal and/or human testing. Until such integration is achieved, the impact of continuum mechanics models will remain bounded.

**Acknowledgements:**

This work is supported by the Medical Research Council (grant number MR/T027932/1 and MR/Z504841/1).

**Author contributions:**

Conceptualization: DS, CY. Supervision: DS, CY. Funding: CY. Writing - Original Draft: DS, JL, MQ. Writing - Review & Editing: CY, TC.

**Conflict of interest:**

The authors declare that they have no conflict of interest.